\documentclass[aps,prl,showpacs,twocolumn]{revtex4}
\usepackage{amsmath}
\usepackage{graphicx}
\usepackage{dcolumn}
\usepackage{longtable}

\newcommand{\be}{\begin{equation}}
\newcommand{\ee}{\end{equation}}
\newcommand{\bearray}{\begin{eqnarray}}
\newcommand{\eearray}{\end{eqnarray}}
\newcommand{\bse}{\begin{subequations}}
\newcommand{\ese}{\end{subequations}}

\begin{document}

\title{Three-photon-annihilation contributions to positronium energies at order $m \alpha^7$.}

\author{Gregory S. Adkins}
\email[]{gadkins@fandm.edu}
\author{Minji Kim}
\author{Christian Parsons}
\affiliation{Franklin \& Marshall College, Lancaster, Pennsylvania 17604}
\author{Richard N. Fell}
\affiliation{Brandeis University, Waltham, Massachusetts 01742}

\date{\today}

\begin{abstract}
Positronium spectroscopy ($n=1$ hyperfine splitting, $n=2$ fine structure, and the 1S-2S interval) has reached a precision of order $1 MHz$.  Vigorous ongoing efforts to improve the experimental results motivate the calculation of the positronium energy levels at order $m \alpha^7$.  In this work we present the result for a complete class of such contributions--those involving virtual annihilation of positronium to three photons in an intermediate state.  We find an energy shift of $2.6216(11) m \alpha^7/(n \pi)^3$ for the spin-triplet $S$ state with principal quantum number $n$.
\end{abstract}


\maketitle


Positronium, the bound state of an electron and its antimatter partner, the positron, is of fundamental interest for a number of reasons.  It represents the purest example of binding in quantum field theory as the constituents are structureless point particles and their low mass implies that the dynamics is dominated by QED--strong and weak interactions effects are negligible.  As a particle-antiparticle bound state, positronium exists in eigenstates of the discrete symmetries parity and charge conjugation, leading to the possibility of leptonic tests of the corresponding symmetries and also to the presence of real and virtual annihilation into photons.  An additional difference between positronium and other bound states such as hydrogen and muonium is the fact that the $m/M$ recoil effects in positronium are maximal due to the equality of electron and positron masses.  Crucially, positronium is accessible to high-precision measurements of its spectrum, decay rates, and branching ratios, and so provides an ideal laboratory for testing fundamental theory and calculational methods for bound state QED.  Confirmed discrepancies between theory and measurement could be a signal for new physics in the leptonic sector.

Several measurements of the positronium $n=1$ and $n=2$ transition frequencies have been done with uncertainties of order $1 MHz$.  These include the $n=1$ hyperfine splitting ($1^3S_1-1^1S_0$), the 1S-2S interval ($2^3S_1-1^3S_1$), and a number of purely $n=2$ intervals \cite{SpecNote}.  For example, the $n=1$ hyperfine (hfs) results with the highest precision are \cite{Mills75,Mills83,Ritter84,Ishida14}
\bearray \label{hfs_expts}
\Delta E_\mathrm{hfs} (\mathrm{Brandeis}) &=& 203\,387.5(1.6)\,MHz \cr
\Delta E_\mathrm{hfs} (\mathrm{Yale}) &=& 203\,389.10(0.74)\,MHz \\
\Delta E_\mathrm{hfs} (\mathrm{Tokyo}) &=& 203\,394.2(1.6)_\mathrm{stat}(1.3)_\mathrm{sys}\,MHz . \nonumber
\eearray
The best result for the 1S-2S interval is \cite{Fee93}
\be
\Delta E(2^3S_1-1^3S_1) = 1\,233\,607\,216.4(3.2)\,MHz.
\ee

There is currently a significant push to explore new and improved approaches to the hfs measurement \cite{Fan96,Sasaki11,Ishida12,Yamazaki12,Namba12,Cassidy12,Miyazaki15} and to improve the 1S-2S result \cite{Cassidy08,Crivelli11,Mills14,Cooke15}.  At the moment, the hfs measurement is the more precise, but there appears to be more potential for improvement in the 1S-2S measurement.  The $\sim \! 1MHz$ uncertainty in the hfs number represents a line-splitting factor of better than one part in a thousand; if a similar splitting of the 1S-2S transition could be achieved (natural line width $\sim 1.3 MHz$), an ultimate experimental precision of a few $kHz$ might be in reach.  Significant improvements in the $n=2$ fine structure \cite{Mills75b,Hatamian87,Hagena93,Ley94,Ley02} would also be possible if a similar line splitting could be performed.

On the theoretical side, all energy contributions through terms of $O(m \alpha^6)$ are known analytically \cite{Elkhovsky94,Pachucki98,Czarnecki99a,Zatorski08}.  In addition, the leading log terms of $O(m \alpha^7 \ln^2 (1/\alpha))$ are known \cite{Karshenboim93,Melnikov99,Pachucki99}, as well as the subleading log contribution to the hfs \cite{Kniehl00,Melnikov01,Hill01}, and a number of pure $O(m \alpha^7)$ contributions \cite{Marcu11,Baker14,Adkins14a,Eides1415,Adkins14b,Adkins15}.  The present theoretical result for the hfs is consistent with the latest of the hfs experiments (``Tokyo'' in Eq.~(\ref{hfs_expts})) but is $\sim 3 \sigma$ above the earlier results.  The theoretical result for the 1S-2S interval is in marginal agreement ($2.4 \sigma$) with the experimental value.

\begin{table}[b]
\begin{center}
\caption{\label{table1} Known contributions to the average 1S energy and $n=1$ hfs at various orders in the perturbative series.  The first column shows the orders that contribute, starting at $m \alpha^4$.  (The leading contribution to the average energy $-m \alpha^2/4$ is not shown.)  The second column gives the numerical value of $m \alpha^4$, $m \alpha^5 \ln(1/\alpha)$, etc.  (The energies are expressed in $MHz$ using the 2014 CODATA recommended values for $R_\infty c$ and $\alpha$.)}
\begin{ruledtabular}
\begin{tabular}{cccc}
order & value & $E_{\mathrm{avg}}$ ($1S$) & $E_{\mathrm{hfs}}$ ($1^3S_1-1^1S_0$) \\
\hline\noalign{\smallskip}
$m \alpha^4$ &350\,377 & 38\,322.493 & 204\,386.630 \\
$m \alpha^5 \ln (1/\alpha)$ &12\,580 & 3\,003.302 & 0 \\
$m \alpha^5$ & 2\,557 & -1\,018.784 & -1\,005.497 \\
$m \alpha^6 \ln (1/\alpha)$ & 91.8 & 2.869 & 19.125 \\
$m \alpha^6$ & 18.7 & 3.000 & -7.330 \\
$m \alpha^7 \ln^2 (1/\alpha)$ & 3.30 & -1.091 & -0.918 \\
$m \alpha^7 \ln (1/\alpha)$ & 0.67 & -- & -0.323 \\
$m \alpha^7$ & 0.14 & -- & -- \\
\end{tabular}
\end{ruledtabular}
\end{center}
\end{table}

The known contributions to the positronium $n=1$ energy levels are shown in Table~\ref{table1}, listed by order in $\alpha$ and $\ln(1/\alpha)$.  Also shown are the numerical values, in $MHz$, of $m \alpha^4$, $m \alpha^5 \ln (1/\alpha)$, etc.  One can see that the actual energy contributions are somewhat smaller, by a factor of $\sim \! 3$, than the values $m \alpha^x \ln^y (1/\alpha)$ of the associated order.  However, pure $m \alpha^7$ contributions as large as several tenths of a $MHz$ have been shown to arise when ``ultrasoft'' photons (energy and momentum $\sim \! m \alpha^2$) are involved \cite{Marcu11,Baker14}.  Knowledge of the full $m \alpha^7$ result is essential for the unambiguous interpretation of present, and certainly future, measurements of the positronium transition energies, especially for the 1S-2S interval.

Positronium energy contributions can be classified by whether or not they involve a complete annihilation of the electron-positron pair into some number of virtual photons as an intermediate state.  Processes involving annihilation into one, two, three, and four photons all contribute at $O(m \alpha^7)$.  Charge conjugation symmetry implies that the $\ell=0$ spin-triplet states $n^3 S_1$ are only affected by one- or three-photon intermediate states.  The one-photon-annihilation contribution at $O(m \alpha^7)$ is known \cite{Baker14}.  The purpose of this letter is to present the result for the $O(m \alpha^7)$ energy level correction due to all processes involving three-photon intermediate states.

\begin{figure}[b]
\includegraphics[width=2.5in]{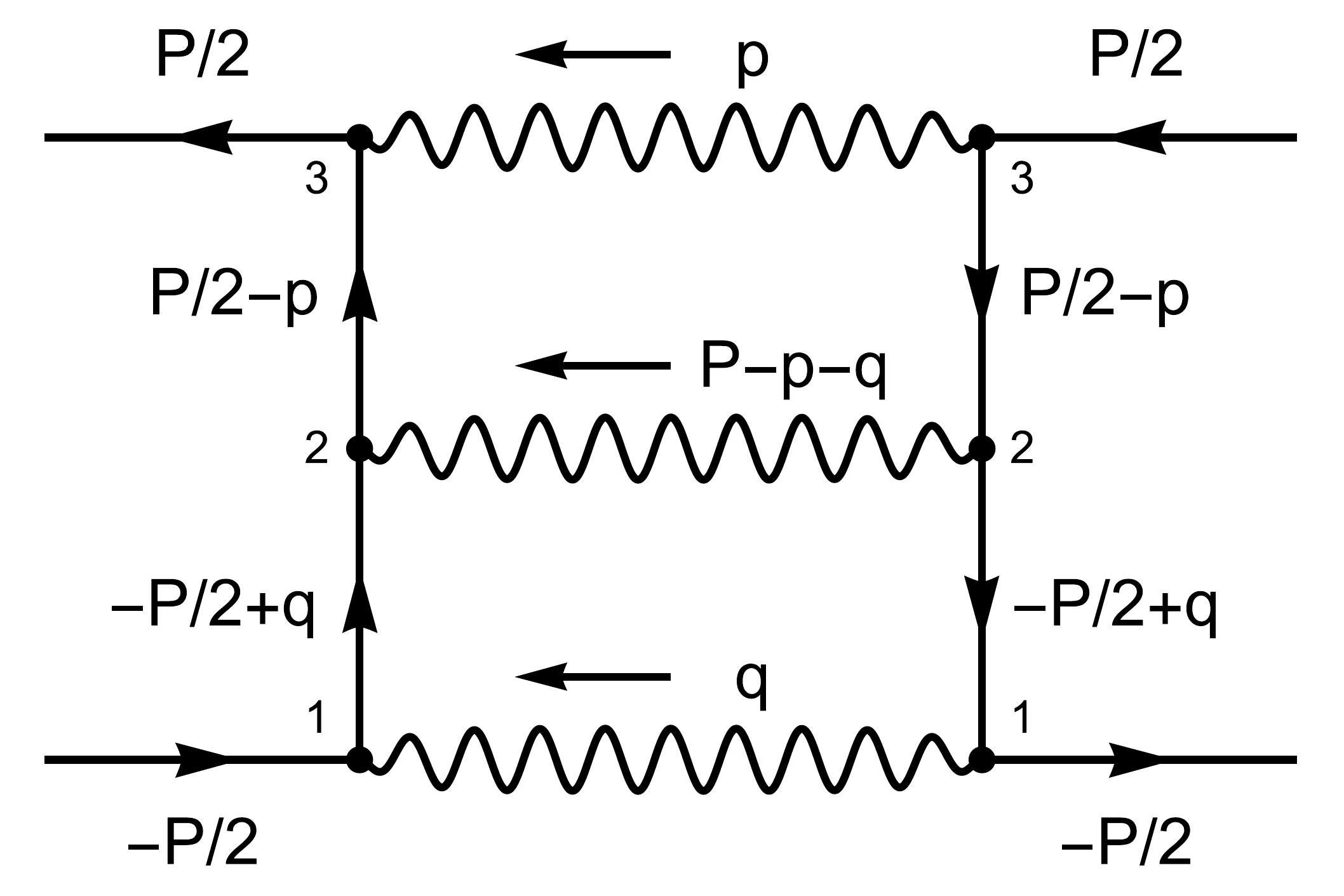}
\caption{\label{fig1} The lowest order $3 \gamma \mathrm{A}$ contribution.}
\end{figure}

The main challenge in dealing with the three-photon-annihilation ($3 \gamma \mathrm{A}$) graphs is the separation of real and imaginary parts.  Our approach is to deal with the graphs non-covariantly by first integrating over the energy components of the annihilation photon loop momenta.  We then identify the terms that contribute imaginary energies by Cutkosky analysis \cite{Cutkosky60} and isolate the imaginary parts as coming from $\ln(-1-i \epsilon)=-i \pi$.  We illustrate this procedure in a calculation of the lowest order $3 \gamma \mathrm{A}$ graph shown in Fig.~\ref{fig1}.  The diagram shown represents one of the six permutations of internal photons that contributes at this order.

The energy shift through $O(m \alpha^7)$ due to $3 \gamma \mathrm{A}$ graphs depends on the spin of the positronium state and the value of the wave function at spatial contact $\phi_0$--other features of the wave function do not enter.  All internal momenta are ``hard''--of $O(m)$--while momenta of order $m \alpha$ and $m \alpha^2$ do not contribute.  Consequently, any convenient bound state formalism can be used.  We employ the formalism of Ref.~\cite{Adkins99}, in which the energy shift is an expectation value
\be \Delta E = i \bar{\Psi} \delta K \Psi .
\ee
The two-body positronium states are, in this approximation,
\be
\Psi \rightarrow \phi_0 \begin{pmatrix} 0 & \chi \\ 0 & 0 \end{pmatrix}
\quad \bar \Psi^T \rightarrow \phi_0 \begin{pmatrix} 0 & 0 \\ \chi^\dagger & 0 \end{pmatrix} ,
\ee
where $\chi$ is the two-by-two two-particle triplet spin matrix $\chi = \vec \sigma \cdot \hat \epsilon/\sqrt{2}$ with $\hat \epsilon$ the orthopositronium polarization.  The  relative momentum vanishes in this approximation, and $\phi_0=\sqrt{m^3 \alpha^3/(8 \pi n^3)}$ is the wave function at spatial contact for a state of principal quantum number $n$ and orbital angular momentum $\ell=0$.

The explicit expression for the energy contribution shown in Fig.~\ref{fig1} is
\bearray \label{energy_LO_explicit}
\Delta E &=& (-1) i \phi_0^2 \int \frac{d^4 p}{(2 \pi)^4} \frac{d^4 q}{(2 \pi)^4} \frac{-i}{p^2} \frac{-i}{(P-p-q)^2} \frac{-i}{q^2} \nonumber \\
&\times& \mathrm{tr} \Bigl [ \begin{pmatrix} 0 & 0 \\ \chi^\dagger & 0 \end{pmatrix} (-i e \gamma^{\mu_3}) \frac{i}{\gamma (P/2-p)-m} \nonumber \\
&\times& \quad  (-i e \gamma^{\mu_2})  \frac{i}{\gamma (-P/2+q)-m} (-i e \gamma^{\mu_1})   \Bigr ] \cr
&\times& \mathrm{tr} \Bigl [ (-i e \gamma_{\mu_1}) \frac{i}{\gamma (-P/2+q)-m} (-i e \gamma_{\mu_2}) \nonumber \\
&\times& \quad \frac{i}{\gamma (P/2-p)-m} (-i e \gamma_{\mu_3}) \begin{pmatrix} 0 & \chi \\ 0 & 0 \end{pmatrix} \Bigr ]
\eearray
where $P=(2m,\vec 0 \,)$ is the positronium 4-momentum in the center-of-mass frame, $p$ and $q$ are the 4-momenta of two of the virtual photons, and the initial $(-1)$ is a fermionic minus sign.  We see that $\Delta E$ is proportional to $m \alpha^6$ times a dimensionless quantity obtained by scaling a factor of $m$ out of each momentum.  The energy shift can be written as
\be \label{energy_1a}
\Delta E = \frac{m \alpha^6}{\pi^2} \frac{2}{3} \int dp dq \, p^2 q^2 \int_{-1}^1 du \int \frac{dp_0}{2 \pi i} \frac{dq_0}{2 \pi i} \frac{T_{1a}}{D_{1a}}
\ee
where $p$ and $q$ now stand for the magnitudes of the dimensionless 3-vectors $\vec p$ and $\vec q$, $u=\hat p \cdot \hat q$, $T_{1a}$ is a product of two traces and the denominator factor is $D_{1a}=(p^2+i \epsilon_1) ((2n-p-q)^2+i \epsilon_2) (q^2+i \epsilon_3) ((p-n)^2-1+i \epsilon_4)^2 ((q-n)^2-1+i \epsilon_5)^2$ where $n=(1,\vec 0 \,)$.  We have used distinct values for the various positive infinitesimals $\epsilon_i$ to facilitate evaluation of the energy integrals, which were performed as contour integrals.  We extended the $p_0$ and $q_0$ contours by appending infinite semicircles in either the upper or lower half planes on which the integrand vanished.  The closed contour integrals were then evaluated by use of the residue theorem.

After performing the $p_0$ and $q_0$ integrals we obtained a number of terms--those giving rise to the imaginary part all contained a denominator factor $p+q+s-2-i \epsilon$, where $s \equiv \vert \vec p + \vec q \, \vert$ and $\epsilon$ here is a positive infinitesimal formed from $\epsilon_1$, $\epsilon_2$, etc.  We isolated the singularity by writing
\be
\frac{1}{p+q+s-2-i \epsilon} = \frac{p+q-s-2}{-2 p q (u-\overline u - i \epsilon)}
\ee
where $\overline u$ is the special value of $u$ for which $p+q+s=2$:
\be
\overline u = -1 + \frac{2(1-p)(1-q)}{p q} = 1 - \frac{2(p+q-1)}{p q} .
\ee
The quantity $\overline u$ has physical values $-1 \le \overline u \le 1$ when $p \le 1$, $q \le 1$, and $p+q \ge 1$.  This region, the orthopositronium ``decay triangle'', corresponds to physical (positive energy) photons with total energy equal to that of positronium ($2m$ in this approximation).  The $u$ integral was written as
\bearray
\int_{-1}^1 du \frac{f(p,q,u)}{u-\overline{u}-i \epsilon} &=& \int_{-1}^1 du \frac{f(p,q,u)-f(p,q,\overline{u})}{u-\overline{u}} \nonumber \\
&+& f(p,q,\overline{u}) \int_{-1}^1 du \frac{1}{u-\overline{u}-i \epsilon},
\eearray
where the final $u$ integral evaluates to $\ln((1-\overline{u})/(1+\overline{u})) + i \pi$ and contains the imaginary part.

The real part of the lowest order $3 \gamma \mathrm{A}$ energy shift $\Delta E_{LO}$ has contributions from the full $0 \le p, q \le \infty$ quadrant.  The imaginary part of $\Delta E_{LO}$ is a 2-dimensional integral and comes only from the decay triangle.  There was no infrared difficulty arising from the loop integrals involving annihilation photons because all six permutations of the annihilation photons were considered together.  We preformed the numerical integrations using the adaptive Monte Carlo integration routine Vegas \cite{Lepage78}.  Quadruple precision was required to control the delicate cancellations among the many parts of the integrands.  Our total result was
\be \label{3gammaA0}
\Delta E_{LO} = \Bigl \{ -0.5126319(20) - 0.30354919(3) i \Bigr \} \frac{m \alpha^6}{\pi^2} .
\ee
The real part of $\Delta E_{LO}$ is consistent with the earlier evaluation of the $3\gamma A$ contribution to the o-Ps energy at $O(m \alpha^6)$ \cite{Cung77,Adkins88,Devoto90}, and the imaginary part gives the corresponding decay rate through $\Gamma_{LO} = -2 \mathrm{Im} (\Delta E_{LO})$ with a numerical value that agrees with the Ore and Powell result $\Gamma_{LO} = \frac{2}{9 \pi} (\pi^2-9) m \alpha^6$ \cite{Ore49}.  These checks show that our method of calculation is working properly and also provide an independent check of the $3 \gamma A$ energy shift.

\begin{figure}
\includegraphics[width=3.2in]{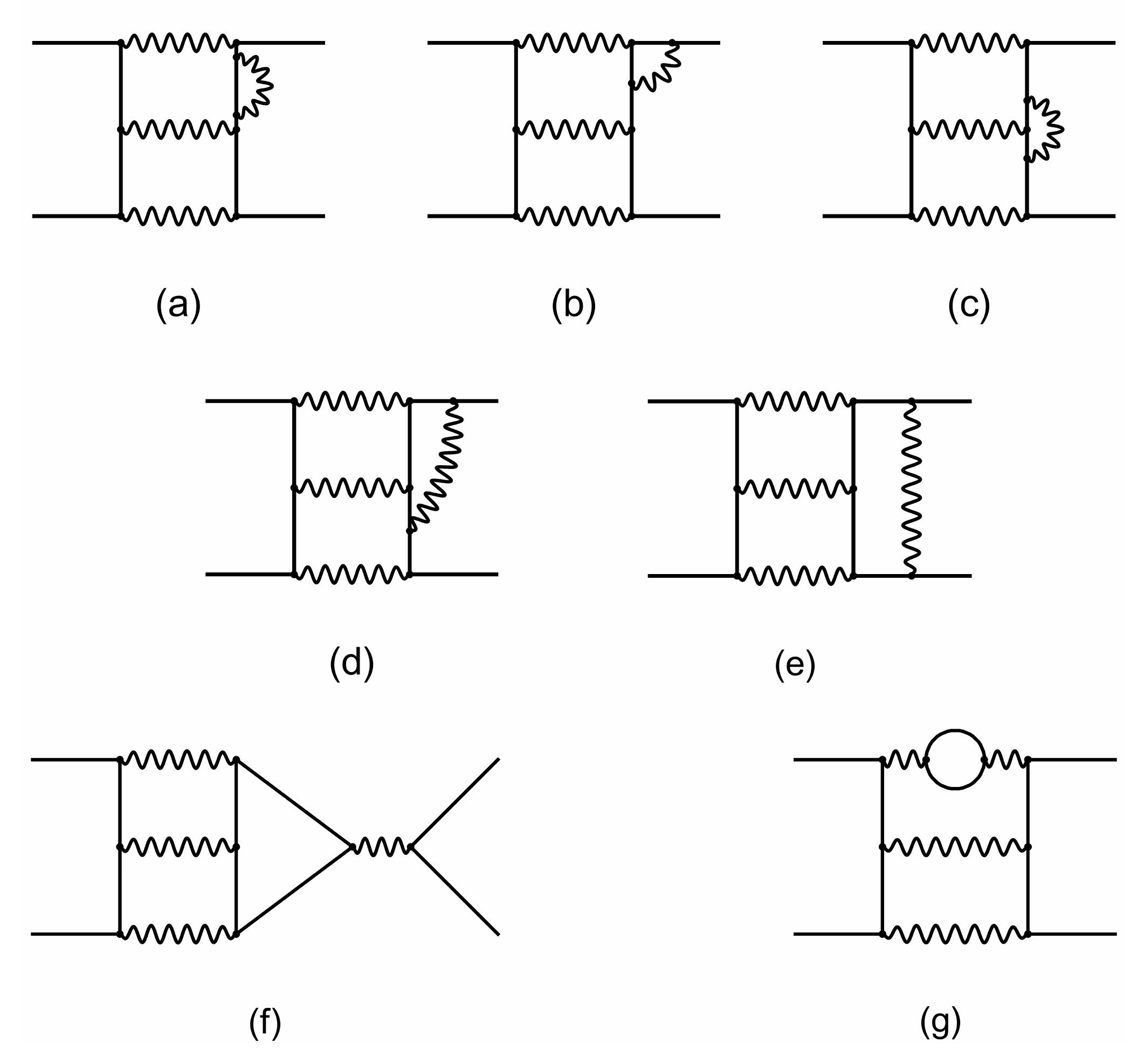}
\caption{\label{fig2} The seven types of one-loop radiative corrections in the $3 \gamma \mathrm{A}$ channel.  Each of these diagrams represents the full set of permutations of the annihilation photons and additional contributions coming from the various places where the correction could act.}
\end{figure}

The one-loop corrections to the lowest order $3\gamma A$ diagram are shown in Fig.~\ref{fig2}.  These terms all give energy contributions of $O(m \alpha^7)$.  They were evaluated one by one as different techniques were required for the various contributions.  Feynman gauge was used throughout.  The self energy (Fig.~\ref{fig2}a) and outer vertex (Fig.~\ref{fig2}b) were fairly straightforward.  We used known expressions in terms of Feynman parameters for the self energy and vertex functions \cite{Adkins01,Adkins15} and calculated the imaginary parts as integrals over the decay triangle and real parts as integrals over the full $pq$ quadrant.  The results are shown in Table~\ref{table2}.  A photon mass $\lambda$ was introduced to allow for mass-shell renormalization of the self energy and vertex parts and to regulate the binding singularity in the ladder graph.  All dependence on $\lambda$ vanished in the net contribution.

The inner vertex contribution of Fig.~\ref{fig2}c was not straightforward.  The parametric expression for the vertex function contains a denominator factor that is a complicated quadratic function of $p_0$ and $q_0$ that can't be easily integrated using the residue theorem.  Instead, we derived an expression for an altered vertex function having the same UV behavior (to allow renormalization) but with no dependence on $q_0$ in the denominator.  The difference is UV-finite, and we evaluated the integral over the loop momentum $k$ non-covariantly using our residue theorem routine for $d k_0$ while $d^3 k$ was done as part of the overall numerical integration.  We found the region of small $p$ and $q$ to be particularly challenging for our numerical integration routine.  We broke up the integration region into a number of parts near $p=q=0$ and evaluated them individually.  The total integral was the result of an extrapolation.

\begin{table}[b]
\begin{center}
\caption{\label{table2} Numerical results for the various contributions to the energy at $O(m \alpha^7)$ due to one-loop corrections to the $3 \gamma \mathrm{A}$ process.  Contributions to $I$ are shown where $\Delta E = (m \alpha^7/\pi^3) I$.   The IR singular terms are proportional to $I_{LO}$, where $\Delta E_{LO} = (m \alpha^6/\pi^2) I_{LO}$.  The real and ``calculated'' imaginary contributions were obtained here via numerical integration.  The ``analytic'' imaginary results come from prior work in which the o-Ps decay rate contributions were calculated exactly \cite{Stroscio82,Adkins8592} or semi-analytically (in terms of two-dimensional integrals \cite{Adkins9605}).}
\begin{ruledtabular}
\begin{tabular}{ccccc}
Term & $\ln \lambda I_{LO}$ & Real & Imaginary & Imaginary \\
&&& (calculated) & (analytic) \\
\hline\noalign{\smallskip}
SE & 4 &  -4.48169(23) & -1.452474(3) & -1.452478 \\
OV & -4 & 4.47981(26) & 0.312310(1) & 0.312310 \\
IV  & -2 & 2.28614(44) & 0.558326(2) & 0.558325 \\
DV & 0 & -1.35555(59) & 1.082948(4) & 1.082944 \\
LAD & 2 & 1.22037(48) & 2.374291(7) & 2.374287 \\
LbyL & 0 & 1.05835(39) & 0.247109(8) & 0.247106 \\
VP & 0 &  -0.585790(2) & 0 & 0 \\
\hline
total & 0 & 2.62164(103) & 3.122510(12) & 3.122494 \\
\end{tabular}
\end{ruledtabular}
\end{center}
\end{table}

The double vertex (Fig.~\ref{fig2}d) was straightforward with no renormalization nor IR regularization required.  We performed the loop integral $d^4 k$ as $d k_0$ using the residue theorem routine and $d^3 k$ numerically.  Again an extrapolation was necessary to achieve an acceptable result for small $p$ and $q$.

The ladder graph (Fig.~\ref{fig2}e) is UV safe but contains a binding singularity that required special treatment.  The IR divergence is regulated by the non-zero photon mass we have assumed throughout.  We separated off the most singular part: $2 I_B (\alpha/\pi) \Delta E_{LO}$ where the binding integral
\be
I_B = \int \frac{d^4 k}{i \pi^2} \frac{-1}{D(k)} = \frac{\pi}{\lambda} + \ln \lambda - 1 + O(\lambda)
\ee
with $D(k) = (k^2-\lambda^2) ((k+n)^2-1) ((k-n)^2-1)$. The $\pi/\lambda$ term is the binding or threshold singularity in this approach to the problem.  Its significance is that the ladder correction contains the basic Coulomb binding that holds the atom together in the first place.  The $\pi/\lambda$ must be discarded, as retaining it would amount to double counting \cite{Caswell77}.  To make the point in another way, exactly this $\pi/\lambda$ would cancel in the matching procedure if the calculation had been done using an effective non relativistic field theory formalism \cite{Caswell86,Adkins02}.  The $(\ln \lambda - 1)$ multiplies an expression proportional to the lowest order $3 \gamma A$ contribution (times $\alpha/\pi$).  The remainder of the ladder contribution is IR finite and was dealt with in the usual way.

The light-by-light graph (Fig.~\ref{fig2}f) promised to be a challenge.  We could not just integrate the loop momentum non-covariantly using our residue theorem routine because the light-by-light loop has a UV divergence as seen by naive power counting (although this divergence vanishes when all permutations are included).  Furthermore, the graph as a whole has an apparent UV divergent by the same power counting argument.  On the other hand, a parametric evaluation of the light-by-light loop (to facilitate implementation of the UV cancellation) seems out of bounds because $p_0$ and $q_0$ would get hopelessly mixed up, while a parametric evaluation of the graph as a whole would give an expression from which the real and imaginary parts would not be easily disentangled.  The key to the evaluation of this graph is the gauge identity satisfied by the light-by-light tensor \cite{Aldins70}
\be \label{gauge_identity}
\Pi_{{\mu_1},{\mu_2},{\mu_3},\lambda} = -P^\kappa \frac{\partial}{\partial P^\lambda} \Pi_{{\mu_1},{\mu_2},{\mu_3},\kappa}
\ee
where $P$ is the 4-momentum of the incoming photon and $\kappa$ is the corresponding Lorentz index.  The gauge identity is obtained by differentiating the condition of gauge invariance as applied to the light-by-light tensor $P^\kappa \Pi_{{\mu_1},{\mu_2},{\mu_3},\kappa}=0$.  The terms on the right hand side of (\ref{gauge_identity}) are each well-behaved in the UV and were treated by non-covariant integration of the fermion loop momentum.

While the light-by-light graph contains a virtual annihilation to a single photon, it was not included in the earlier evaluation \cite{Baker14} of one-photon-annihilation contributions \cite{Penin_7_13_2015}.

The vacuum polarization contribution (Fig.~\ref{fig2}g) is purely real because the vacuum polarization--corrected photon does not have the usual pole at $p^2=0$.  We used a standard parametrization of the vacuum polarization function \cite{Adkins15} and performed the integrals over $p$ and $q$ in two ways: by Feynman parameters and non-covariantly using our residue theorem routine.  The results of the two approaches were consistent.

The results for the various contributions are summarized in Table~\ref{table2}.  The net IR divergence vanishes, as it must.  The imaginary part of each contribution is in accord with the corresponding result of prior work.  The net result of all $O(\alpha)$ corrections in the three-photon-annihilation channel is
\be \label{3gammaA1}
\Delta E = \left \{ 2.6216(11) + 3.122510(12) \, i \right \} \frac{m \alpha^7}{\pi^3} .
\ee

The energy level shift implied by this correction is
\be
\Delta E = 2.6216(11) \frac{m \alpha^7}{\pi^3} \frac{\delta_{\ell,0} \delta_{s,1}}{n^3} .
\ee
for states with principal quantum number $n$, orbital angular momentum $\ell$, and total spin $s$.  For o-Ps in the ground state this amounts to $11.5 kHz$.  On comparing (\ref{3gammaA0}) and (\ref{3gammaA1}) we see that these corrections involve sizable factors:
\begin{subequations}
\bearray 
\mathrm{Re} \Delta E_{3\gamma A} &=& -0.5126 \Bigl \{ 1-5.114(2) \frac{\alpha}{\pi} \Bigr \} \frac{m \alpha^6}{\pi^2} ,\\
\mathrm{Im} \Delta E_{3 \gamma A} &=& -0.3035 \Bigl \{ 1-10.2866 \frac{\alpha}{\pi} \Bigr \} \frac{m \alpha^6}{\pi^2} .
\eearray
\end{subequations}
The correction to the imaginary part is easily seen in o-Ps lifetime measurements \cite{Kataoka09}.  The $3 \gamma \mathrm{A}$ energy correction is relevant to all classes of positronium spectroscopy: 1S-2S, hfs, and $n=2$ fine structure.  It should certainly be seen if suggested improvements in the 1S-2S measurements can be realized.


\begin{acknowledgments}
We are grateful to Zvi Bern for an illuminating discussion about the analysis of complicated diagrams.  We thank Jason Brooks and Anthony Weaver for useful suggestions on numerical integration.  We acknowledge the support of the National Science Foundation through Grant No. PHY-1404268.
\end{acknowledgments}


     


\end{document}